\documentclass[aps,prb,floatfix,twocolumn,tightenlines,amsmath,amssymb]{revtex4}

\usepackage{graphicx}
\usepackage{amsmath}
\usepackage{amssymb}
\usepackage{epstopdf}
\usepackage{color}

\newcommand{\ket}[1] {|#1 \rangle}

\newcommand*{\rom}[1]{\expandafter\@slowromancap\romannumeral #1@}

\begin{document}

\title{Extracting inter-dot tunnel couplings between few donor quantum dots in silicon}
\author{S. K. Gorman, M. A. Broome, J. G. Keizer, T. F. Watson, S. J. Hile, W. J. Baker, M. Y. Simmons}
\affiliation{Centre of Excellence for Quantum Computation and Communication Technology, School of Physics, University of New South Wales, Sydney, New South Wales 2052, Australia}
\date{\today}

\begin{abstract}
The long term scaling prospects for solid-state quantum computing architectures relies heavily on the ability to simply and reliably measure and control the coherent electron interaction strength, known as the tunnel coupling, $t_c$. Here, we describe a method to extract the $t_c$ between two quantum dots (QDs) utilising their different tunnel rates to a reservoir. We demonstrate the technique on a few donor triple QD tunnel coupled to a nearby single-electron transistor (SET) in silicon. The device was patterned using scanning tunneling microscopy-hydrogen lithography allowing for a direct measurement of the tunnel coupling for a given inter-dot distance. We extract $t_c{=}5.5{\pm}1.8$~GHz and $t_c{=}2.2{\pm}1.3$~GHz between each of the nearest-neighbour QDs which are separated by 14.5 nm and 14.0 nm, respectively. The technique allows for an accurate measurement of $t_c$ for nanoscale devices even when it is smaller than the electron temperature and is an ideal characterisation tool for multi-dot systems with a charge sensor.
\end{abstract}

\maketitle

The entanglement of multiple quantum particles is becoming an established practice for enhanced measurement protocols in quantum metrology~\cite{giovannetti2011}, secure communications in quantum key distribution~\cite{hwang2003,lo1999} and is the central tenant of quantum computation~\cite{divincenzo2000a}. Entanglement is created by a coherent coupling between quantum particles. In solid-state architectures, the spin-spin interaction between single electrons isolated to quantum dots (QDs) enables multi-qubit operations needed for universal quantum computation~\cite{levy2002}. The strength of this interaction is governed by the coherent tunnel coupling term, $t_c$ between two electron charge states of neighbouring QDs~\cite{takakura2014}.

Unlike in conventional QD architectures where electrons are confined using metallic surface gates, donor based systems rely on the attractive Coulomb potential of the donor atoms themselves~\cite{kane1998,pierre2009,fuechsle2010,mazzeo2012,buch2013}. Nanoelectronic devices based on phosphorus doped silicon (Si:P) have recently demonstrated electron transport at the few electron level where spin-spin interactions can be observed~\cite{roche2012,weber2014,watson2014}. Following this, the singlet-triplet states of a strongly coupled donor pair have been readout~\cite{dehollain2014} and electrons confined to double QDs formed from donor clusters have been investigated using charge-sensing~\cite{watson2015b}. Importantly, for the scalability of multi-donor systems in a quantum computing architecture, the ability to simply and reliably measure $t_c$ between donors is vital.

Unlike gate defined QDs~\cite{dicarlo2004,petta2004,koppens2005}, the value of $t_c$ between isolated phosphorus donors or clusters in Si:P qubit architectures is fixed by the physical distance between the donors~\cite{koiller2002,wellard2003,hollenberg2006}, and is difficult to tune using external gates since the donors are only separated by tens of nanometres~\cite{weber2014,kane1998,skinner2003}. Therefore, knowledge of $t_c$ as a function of donor separation is extremely important for the design, fabrication, and operation of donor based qubits~\cite{koiller2002,koiller2002b,wellard2003,hollenberg2006}. Several methods to determine $t_c$ based on electron transport~\cite{nazarov1993,waugh1996,vanderwiel2003,dupont-ferrier2013}, spin funnel experiments~\cite{petta2005,maune2012} and the response of a quantum point contact across an inter-dot charge transition~\cite{dicarlo2004} have already been established. However, these techniques either require multiple-electron spin readout at low magnetic fields~\cite{petta2005} or a large capacitive difference between the QDs and charge sensor~\cite{dicarlo2004}. The second condition requires that both QDs are at vastly different distances to the charge sensor; which is not ideal for single electron spin readout since complex shuttling protocols must be developed to determine the individual spin states.

In this letter we demonstrate a new method to determine $t_c$ between QDs based on a simple time-resolved charge sensing technique. We use our method to determine $t_c$ between adjacent QDs in a triple QD device that uses a single-electron transistor (SET) as a charge sensor. SETs have been used extensively for charge sensing in Si:P where they can perform high-fidelity single-electron spin readout~\cite{morello2010,pla2012,buch2013,watson2015}. We show how our time-resolved SET charge sensing technique can be used to determine $t_c$ and at the same time allows for the extraction of electron temperature, $T_e$. The gate lever arms, $\alpha$, can also be measured without the possibility of artificial broadening of the SET Coulomb blockade peaks~\cite{kung2012} present in previously reported methods~\cite{morello2010}.

\begin{figure}
\begin{center}
\includegraphics[width=1\columnwidth]{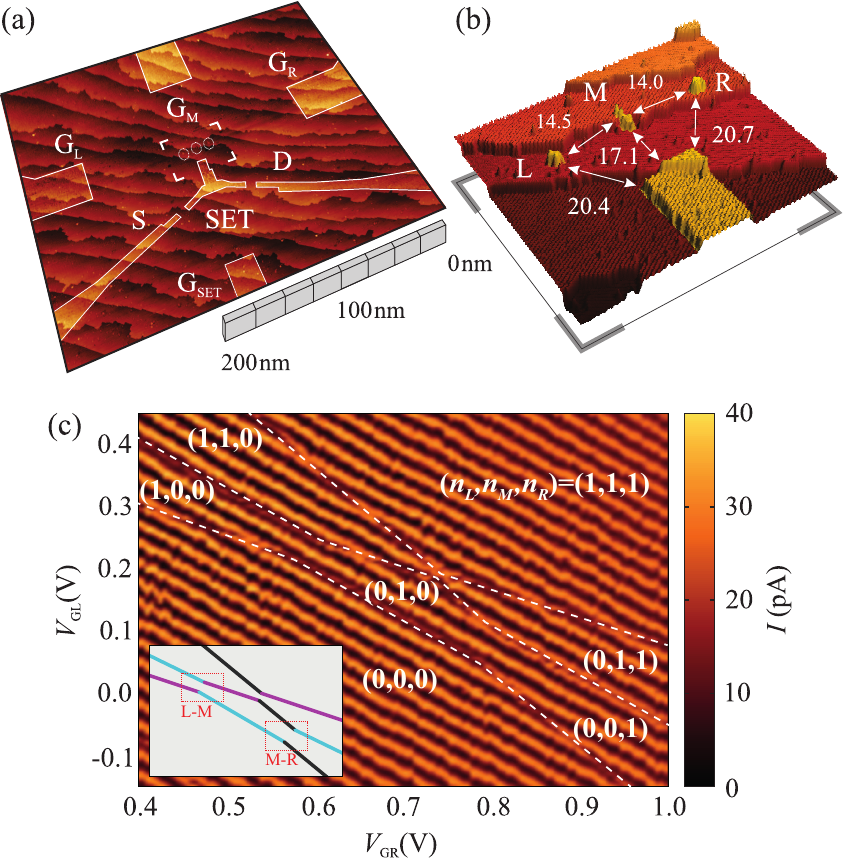}
\end{center}
\caption{A ${\sim}5$ donor triple quantum dot in Si:P with an adjacent singlet-electron transistor used as a charge sensor (all distances are in nm). (a) A scanning tunneling micrograph showing the lithographic outline of the device before PH$_3$ dosing. The device consists of four control gates (G$_L$, G$_M$, G$_R$, G$_{SET}$), a SET with source (S) and drain (D) leads, and three small QDs tunnel coupled to the SET. (b) A close-up of the three QDs. From the lithographic area it is estimated that they will contain ${\sim}5$ P donors. Neighbouring QDs are separated by $14.5$ nm and $14.0$ nm for the left-middle and middle-right, respectively. (c) The current through the SET as a function of the voltage on $G_L$ and $G_R$. The breaks in the SET Coulomb blockade peaks show where the QD electrochemical potentials align with the SET and hence where a charge transition of the QD will occur. By following the SET breaks the typical pentagon structure of a triple QD can be seen, where the equivalent charge numbers are given by ($n_L,n_M,n_R$). The inset shows a schematic of the quadruple point with two dotted red boxed indicating the inter-dot transitions investigated in this work.}
\label{fig:intro}
\end{figure}

The device, shown in Fig.~\ref{fig:intro}, was fabricated using a scanning tunneling microscope (STM) to selectively remove hydrogen from a passivated Si(100) $2{\times}1$ reconstructed surface. The lithographic mask is subsequently dosed with PH$_3$ and annealed to incorporate P atoms into the silicon substrate~\cite{fuhrer2009}. The lithographic outline of the device is shown in Fig.~\ref{fig:intro}a,b. The QDs, $L$, $M$, and $R$ (left, middle, and right) are formed with ${\sim}5$ P donors in each QD, determined by examining the size of the lithographic patches~\cite{buch2013,weber2014}. Three gates, G$_L$, G$_M$ and G$_R$ control the electron numbers on the QDs. The electrons are able to tunnel to a SET island used as the charge sensor. The SET is defined as a much larger QD in between source and drain leads with a control gate G$_{SET}$. It is operated with a source-drain bias of 0.3 mV and has a charging energy of ${\sim}5$~meV.

In our experiment, G$_L$ and G$_R$ are used to detune QDs $L$ and $R$ with respect to the SET, while G$_M$ is used as global gate to shift the potential of the QDs. Figure~\ref{fig:intro}c shows the SET transport current as a function of G$_L$ and G$_R$. Enhanced current lines running at ${\sim}45^{\circ}$ due to Coulomb blockade of the SET can be seen with breaks corresponding to charge transitions of the three QDs. Due to the different capacitive coupling of the gates to the QDs, three lines of SET breaks with distinct slopes are visible. In addition a characteristic pentagon structure associated with the quadruple point of a triple QD~\cite{busl2013,braakman2013,watson2014} can be seen, confirming the presence of three separate QDs. We note that the absolute electron number has not been determined for this device; however, for the purpose of this work we assign the charge states shown in Fig.~\ref{fig:intro}c where ($n_L,n_M,n_R$) represents the relative electron numbers on QDs $L$, $M$ and $R$, respectively.

\begin{figure}
\begin{center}
\includegraphics[width=1\columnwidth]{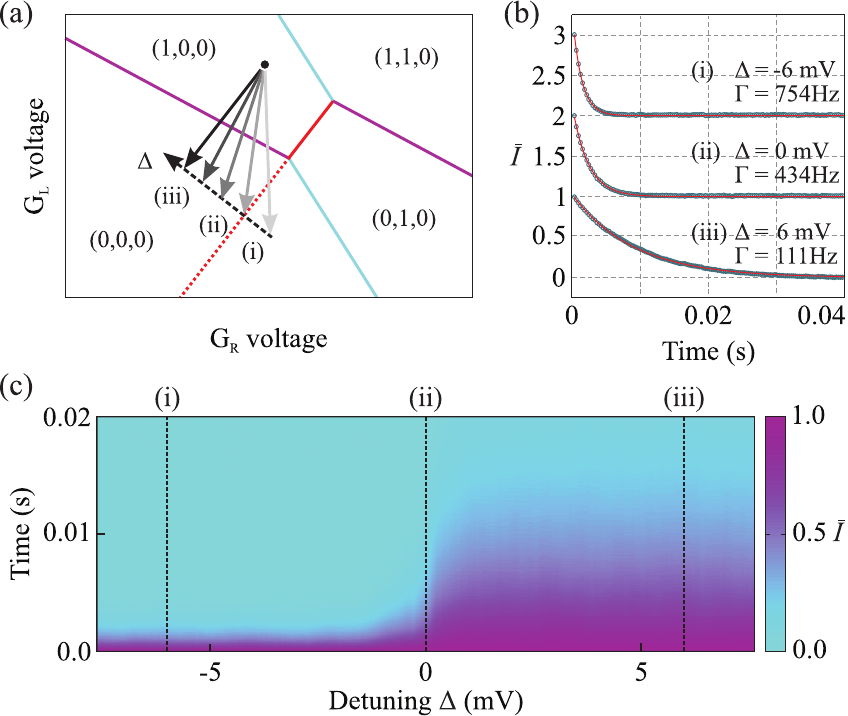}
\end{center}
\caption{Protocol for measurement of the inter-dot tunnel coupling utilising the difference in tunnel rates of the QDs to the SET. (a) A schematic of the $L$-$M$ inter-dot transition showing the gate pulses required to measure the inter-dot tunnel coupling. The measurement protocol begins by loading an electron onto the left QD so that the system is in the (1,0,0) charge state. The gates are then pulsed adiabatically with respect to $t_c$ but faster than the tunnel rate to the SET to move into the (0,0,0) state. The solid arrow shows the start and end point of a two level pulse to load and unload an electron from the (1,0,0)$\rightarrow$(0,0,0). The dashed arrow shows the detuning, $\Delta$ for the unload phase of the two-level pulse; it sweeps past the inter-dot transition, marked by the red dashed line. (b) The normalised SET current, $\bar{I}$ showing the tunnel time, $\Gamma(\Delta)$ for different positions of detuning, $\Delta$ (the curves are offset by 1). (c) The normalised current map across the $L$-$M$ charge transition showing the tunnel rate, $\Gamma(\Delta)$ as a function of $\Delta$ across the inter-dot transition.}
\label{fig:expt}
\end{figure}

Next, we describe the method of determining $t_c$ using the $L$-$M$ transition as an example, see Fig.~\ref{fig:expt}. The protocol involves measuring the tunnel rate from the QDs to the SET across an inter-dot transition, Fig.~\ref{fig:expt}a. From the detuning dependency of the measured tunnel rates, the $t_c$ can be extracted. Using a two level pulse scheme, the system is first initialised in the equivalent single electron state, (1,0,0) after which the second pulse moves into the (0,0,0) to unload this electron. This pulse duration, $t_p = 0.1$ ms, is adiabatic with respect to $t_c$ but faster than either of the independent tunnel rates from the QDs to the SET. An exponential decay is fitted to the average current trace (200 cycles) and a tunneling rate $\Gamma(\Delta)$ extracted accordingly, see Fig.~\ref{fig:expt}b. The protocol involves varying the unload point along a detuning axis, $\Delta$ across the inter-dot transition $L$-$M$ and measuring the SET current $I_{SET}(t)$ as a function of time, shown in Fig.~\ref{fig:expt}c. The procedure for the $M$-$R$ transition is the same; however, the initial charge configuration is chosen to be (0,0,1).

Far from the inter-dot transition at points (i) and (iii) in Fig.~\ref{fig:expt}a the tunnel rate $\gamma_L{=}\Gamma(\Delta{=}{-}6\textnormal{ mV}){=}111$~Hz and $\gamma_M{=}\Gamma(\Delta{=}6\textnormal{ mV}){=}754$~Hz, respectively. However, as the unload point moves closer toward the inter-dot transition (shown by the dashed red line in Fig.~\ref{fig:expt}a) $\Gamma(\Delta){\neq}\gamma_L$ or $\gamma_M$ and is given instead by the expression,
\begin{equation}
\Gamma(\Delta) \approx \sum_{i={L,M}}P_i \Gamma_i,
\label{eq:sumgamma}
\end{equation}
where $P_i$ is the probability of the electron occupying QD $i{=}L,M$ and $\Gamma_i$ is the effective tunnel rate which takes into account assisted tunneling via a neighbouring QD to the SET. Importantly both the occupation probabilities and effective tunnel rates will depend on the parameters, $\Delta$, $t_c$ and temperature of the system, $T$~\cite{dicarlo2004}. We have also performed numerical calculations based on a Linblad master equation approach and achieve the same form for the tunnel rate, $\Gamma(\Delta)$ (see appendix).

\begin{figure}
\begin{center}
\includegraphics[width=1\columnwidth]{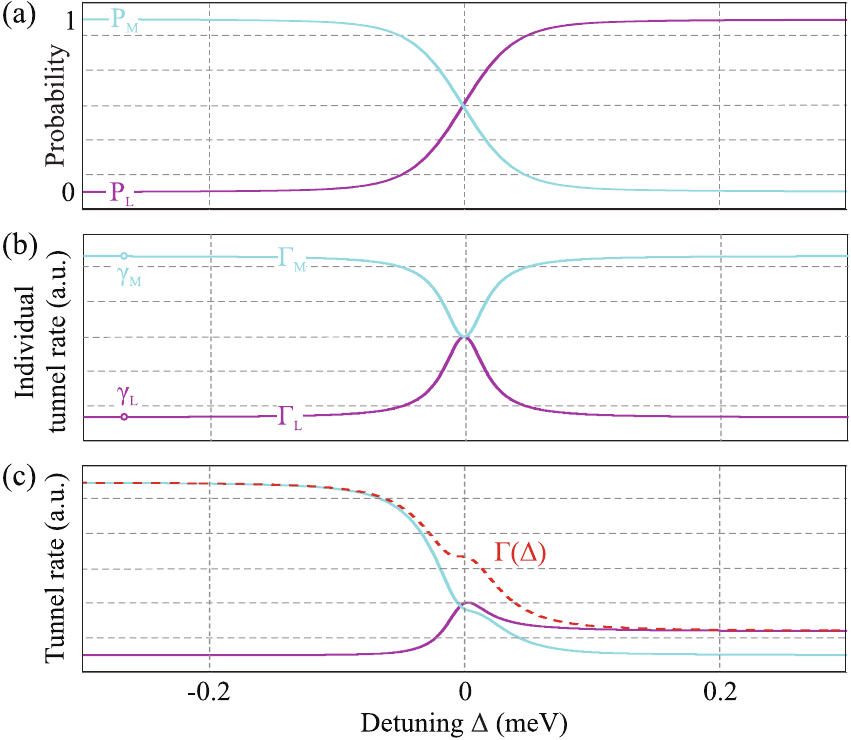}
\end{center}
\caption{Theoretical lineshape of the tunnel rates near the inter-dot transition. (a) The theoretical probability of an electron residing in each QD as a function of $\Delta$ at a temperature of $T{=}200$~mK. (b) The predicted effective tunnel rate to a SET for two tunnel coupled quantum dots (QDs) with $t_c{=}3.5$~GHz. At zero detuning, $\Delta{=}0$ the tunnel rates to the SET must be equal since the electron is delocalised over both QDs. (c) Solid lines show the corresponding tunnel rates from each QD weighted by the electron occupation probability, \textit{i.e.} the product of the traces shown in (a) and (b). The dashed line shows the sum of the weighted tunnel rates, $P_i \Gamma_i$, from Eq.~(\ref{eq:sumgamma}), which theoretically predicts the form of the measured tunnel rate to the SET, $\Gamma(\Delta)$.}
\label{fig:theory}
\end{figure}

To find the general expression of $\Gamma(\Delta)$ we must compute the probabilities $P_i$ of the electrons occupying the QDs in Eq.~(\ref{eq:sumgamma}). Assuming that the QDs are in thermal contact with the SET with a temperature, $T$, since $k_B T \gg \gamma_i$~\cite{house2015}, the probabilities of finding the electron in either of the two QDs is given by~\cite{dicarlo2004},
\begin{equation}
P_{\tiny{\begin{Bmatrix}
L\\
M \end{Bmatrix}}} = \frac{1}{2} \pm \frac{\Delta \tanh{\Big(\frac{\theta}{2 k_B T}\Big)}}{2\theta},
\label{eq:probs}
\end{equation}
where $\theta{=}\sqrt{\Delta^2 + t_c^2}$, see Fig.~\ref{fig:theory}a.

The shape of the (1,0,0)-(0,1,0) anti-crossing as a function of detuning, $\Delta$ dictates that the rates $\Gamma_i$ follow a Lorentzian given by~\cite{watson2015},
\begin{equation}
\Gamma_i = \gamma_i - \frac{\delta\gamma_{i,j} t_c^2}{2(\Delta^2 + t_c^2)},	
\label{eq:rates}
\end{equation}
where $\delta\gamma_{i,j}{=}\gamma_i{-}\gamma_j$, see Fig.~\ref{fig:theory}b. This means that at $\Delta{=}0$ the effective tunnel rates are equal, $\Gamma_L{=}\Gamma_M{=}(\gamma_L{+}\gamma_M){/}2{=}\bar{\gamma}_{LM}$, because the electron is fully delocalised over both QDs. In this analysis we have neglected the many excited states of the SET (that form a quasi-continuum) since the large detuning of the unload position makes the QD chemical potentials much higher than the SET Fermi level. This means the excited state energy levels are small compared to the overall energy scale. In addition, in the experiment we ensure that the unload position always follows the same Coulomb peak of the SET such that the same number of electrons are present on the SET.

Substituting Eq.~(\ref{eq:probs}) and (\ref{eq:rates}) into Eq.~(\ref{eq:sumgamma}) we find,
\begin{equation}
\Gamma(\Delta) \approx \bar{\gamma}_{LM} + \frac{\Delta^3(\delta\gamma_{LM})\tanh{\Big(\frac{\theta}{2 k_B T}\Big)}}{2\theta (\Delta^2 + t_c^2)}.
\label{eq:gamma}
\end{equation}
Figure~\ref{fig:theory} shows the functional forms of $\Gamma_L$ and $\Gamma_M$, $P_L$ and $P_M$, and $\Gamma(\Delta)$ as a function of detuning and for a fixed temperature ($T{=}200$~mK) and tunnel coupling ($t_c{=}3.5$~GHz). In Fig.~\ref{fig:theory}c, the red dashed line corresponding to $\Gamma(\Delta)$ shows a plateau around $\Delta{=}0$, where $\Gamma_L{=}\Gamma_M$. The width of the plateau is directly related to the strength of $t_c$. We can eliminate the possibility that the plateau is caused by an external charge fluctuation by looking at the stability map, Fig.~\ref{fig:intro}c. There is no charge offset in the SET current that can be attributed to external charge movement. Equation~\ref{eq:gamma} diverges from the theoretically predicted results using the Lindblad formalism for $t_c \gg k_B T$ where the sequential tunneling approach cannot be used.

It is worth noting that if $\gamma_L{=}\gamma_M$, then $\Gamma(\Delta)$ does not vary as a function of $\Delta$ and our method cannot be used to gain any information about $t_c$. As a result, $\delta\gamma_{LM}{\neq}0$ is required which is most likely the case for any system, in particular for donor based QDs where the coupling decays exponentially with distance~\cite{koiller2002,wellard2003} meaning differences in donor position even on the atomic scale will change the tunnel rate significantly.

By fitting Eq.~(\ref{eq:gamma}) to the tunnel rate across the transition we can determine $t_c$ and the temperature, $T$. By varying the temperature we are able to determine both the minimum electron temperature, $T_e$ as well as the lever arm for the inter-dot transition, $\alpha$~\cite{morello2010}.

\begin{figure}
\begin{center}
\includegraphics[width=1\columnwidth]{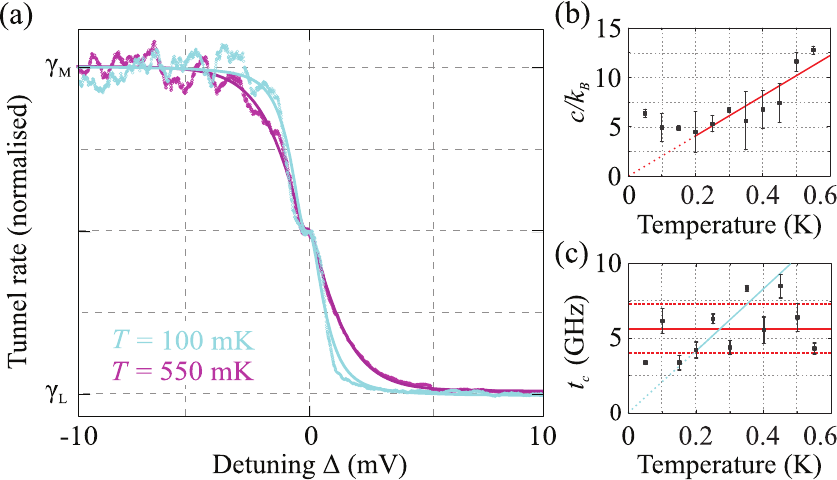}
\end{center}
\caption{Temperature dependence of $\Gamma(\Delta)$ for the lever arm calculation. (a) The temperature broadening effect of the tunnel rate, $\Gamma(\Delta)$ with a fit to Eq.~(\ref{eq:gamma}). The plateau near zero detuning can be clearly seen at both $T{=}100$~mK and $T{=}550$~mK allowing for calculation of $t_c$ at multiple temperatures. (b) The lever arm, $\alpha$ of the inter-dot transition can be found by repeating the  measurement at various temperatures and plotting the fitting parameter, $c/k_B$. A fit to the section where $c/k_B$ is linear (red solid line) gives $\alpha{=}0.0491{\pm}0.0028$ as the inverse slope. The temperature at which $c/k_B$ deviates from the linear fit is taken as the minimum $T_e{\sim}200$ mK. (c) The experimentally determined tunnel coupling, $t_c$ between $L$-$M$ showing no temperature dependence. The solid red line shows the average tunnel coupling, with an error of one standard deviation given by the dashed lines. The blue solid line indicates the thermal energy $k_B T$ demonstrating the measurement of $t_c{<}k_B T$ for $T{\geq}300$ mK.}
\label{fig:temp}
\end{figure}

Figure~\ref{fig:temp}a, shows the form of $\Gamma(\Delta)$ at two different temperatures $T{=}100$~mK and $550$~mK where we see the impact of thermal broadening. The lever arm is related to the temperature, $T$ by,
\begin{equation}
c = \frac{k_B T}{\alpha},
\end{equation}
where $c$ is a fitting parameter used in place of $k_B T$ in Eq.~(\ref{eq:gamma}). By fitting a line to $c / k_B$ as a function of temperature we can extract the lever arm from the inverse of the slope, $1/\alpha$ as in Fig.~\ref{fig:temp}b. For the $L$-$M$ transition we find $\alpha{=}0.0491{\pm}0.0028$ and for the $M$-$R$ transition $\alpha{=}0.0659{\pm}0.0142$. The error in the $M$-$R$ transition $\alpha$ is much larger compared to $L$-$M$ transition $\alpha$ because the temperature dependence was not performed on this transition. Instead, the lever arm was determined by assuming a $T_e{=}200$~mK, calculated from the $L$-$M$ transition temperature dependence in Fig.~\ref{fig:temp}. The temperature at which $c/k_B$ deviates from the linear fit is where $T_e{\neq}T$ and is taken as the minimum $T_e$ of the system, here we find $T_e{\sim}200$~mK.

Finally, from $\Gamma(\Delta)$ and $\alpha$ we can calculate $t_c$. For the $L$-$M$ charge transition, we find $t_c{=}5.5{\pm}1.8$~GHz and for the $M$-$R$ transition $t_c{=}2.2{\pm}1.3$~GHz. The technique allows $t_c$ to be measured even when $k_B T{>}t_c$ since the width of the plateau only depends on $t_c$. Therefore, $t_c$ can also be measured as a function of temperature. The $t_c$ should not have a temperature dependence since it corresponds to an energy separation. This is confirmed experimentally in Fig.~\ref{fig:temp}c where $t_c$ remains constant as the temperature is varied. The coherent coupling term between the QDs is much greater than the inelastic tunnel rate to the SET. This is a complex problem involving the shape of the electron wavefunctions in the SET and QDs, the crystallographic orientation, and the inelastic processes that gives rise to the tunnel event from the QD to the SET. We note that it is difficult to make a direct comparison to previously measured or theoretical results for single donors~\cite{koiller2002,koiller2002b,wellard2003} since both the absolute electron and donor numbers are not known for our device. However, the $t_c$ values obtained from this work are consistent with previously reported values with slightly different donor numbers and inter-dot distances~\cite{weber2014,watson2015b}.

In summary, we have demonstrated a relatively simple method to determine the tunnel coupling between adjacent QDs or donors that are  tunnel coupled to a reservoir. The method can be applied to any system with a charge sensor where tunnel times can be measured and can also be used to obtain the minimum electron temperature and lever arms. The simplicity of the technique makes it ideal as a characterisation tool for future experiments examining multi-donor exchange interactions. Combining this method with atomic-precision lithography using a STM we can investigate the relationship between the inter-dot distance and the exchange coupling predicted from theoretical calculations~\cite{koiller2002,wellard2003,koiller2002b}.

This research was conducted by the Australian Research Council Centre of Excellence for Quantum Computation and Communication Technology (project no. CE110001027) and the US National Security Agency and US Army Research Office (contract no. W911NF-08-1-0527). M.Y.S. acknowledges an Australian Research Council Laureate Fellowship.

\begin{appendix}
\section{Numerical modelling of the tunnel rate across a double quantum dot}

We model the system with a Hubbard-like Hamiltonian given by,
\begin{equation}
H = \sum_{i} (\epsilon_{i} - \mu_i) c^{\dag}_i c_i + \sum_{i \neq j} \frac{t_{c(i,j)}}{2} (c^{\dag}_i c_j + c^{\dag}_j c_i),
\end{equation}
where, $\epsilon_{i}$ and $\mu_i$ are the detuning and chemical potential of QD $i$, $c_i$ ($c_i^{\dag}$) is the annihilation (creation) operator for the electron on QD $i$ and $t_{c(i,j)}$ is the coupling term between the QDs. We assume that only 1 electron can be present in the system and work in the basis $\{\ket{0},\ket{L},\ket{M}\}$. To model the tunnel rates out of the QDs we assume that after the initial pulse from the (1,0,0) state to the unload position into the (0,0,0) the electron finds itself in an eigenstate of $H$ at each value of detuning. This assumption is valid when the $t_{c(i,j)}$ is much greater than the tunnel rate out of either QD. Based on this assumption, the eigenstate of the system in the (0,0,0) will be given by,
\begin{equation}
\rho_{g} = \frac{1}{Z}e^{-H/k_B T},
\end{equation}
with $Z{=}$Tr$(e^{-H/k_B T})$, $k_B$ is Boltzmann's constant and $T$ is the electron temperature. However, since we wait a time $t \gg T_2$ we let the off-diagonal terms in $\rho_g$ go to zero. The inelastic tunnel rates for each QD are included by transforming the system from Hilbert space to Liouville space and assuming the Markov approximation~\cite{gorman2015}. In Liouville space the incoherent contribution is given by,
\begin{equation}
\mathcal{L}_{\Gamma} = \sum_{i} \frac{\Gamma_{i}}{2} (2 c_i \otimes c_i - c_i^{\dag} c_i \otimes \mathbb{I}\\ - \mathbb{I} \otimes c_i^{\dag} c_i),
\end{equation}
Finally, we solve the Lindblad equation as a function of the detuning across the anti-crossing between the two charge states,
\begin{equation}
\dot{\rho} = i(\mathbb{I} \otimes H - H \otimes \mathbb{I}) \rho_g + \mathcal{L}_{\Gamma} \rho_g.
\end{equation}
We watch $\rho_{00}(t)$ during the time evolution. The time dependence of $\rho_{00}(t)$ will follow a double exponential decay due to the tunneling out of the states $\ket{L}$ and $\ket{M}$,
\begin{equation}
\rho_{00}(t) = 1 - \rho_{LL}(0) e^{-t \Gamma_L} - \rho_{MM}(0) e^{-t \Gamma_M},
\end{equation}
where $\rho_{LL}(0)$ and $\rho_{MM}(0)$ are calculated from $\rho_g$. In the experiment we fit the decay to a single exponential, effectively taking a waited average of the double exponential decay. This is the origin of Eq.~\ref{eq:sumgamma} in the main text. We find that the time when $\rho_{00}(t) =  1 - e^{-1}$ agrees well with Eq.~\ref{eq:sumgamma} for most values of $t_c$ and deviates in the limit that $k_B T / t_c \rightarrow 0$. However, the width of the plateau region, which is proportional to $t_c$ remains valid for all values of $k_B T/ t_c$.

\end{appendix}

\bibliography{MyBib}
\bibliographystyle{apsrev}

\end{document}